\documentclass[final,5p,times,twocolumn]{elsarticle}

\usepackage[utf8]{inputenc}
\usepackage{amsmath}
\usepackage{amssymb}
\usepackage{bm}
\usepackage{color}
\usepackage{natbib}
\usepackage{hyperref}

\newcommand{\Pe}{\ensuremath{\text{Pe}_\text{r}}}
\newcommand{\vel}{\mathrm{v}} 

\definecolor{Green}{rgb}{0,0.5,0}

\journal{Current Opinion in Colloid \& Interface Science}

\begin{document}

\begin{frontmatter}


 \title{{Minimal model of active colloids highlights the role of mechanical interactions in controlling the emergent behavior of active matter} }
 \author[label1,label2]{M. Cristina Marchetti}
 \ead{mcmarche@syr.edu}

\author[label4]{{Yaouen Fily}}
\ead{yffily@gmail.com}
 
 \author[label3]{{Silke Henkes}}
\ead{shenkes@abdn.ac.uk}

\author[label1]{Adam Patch}
\ead{apatch@syr.edu}

\author[label1,label5]{David Yllanes} 
\ead{dyllanes@syr.edu}

 \address[label1]{Physics Department, Syracuse University, Syracuse, NY 13244\fnref{label1}}
  \address[label2]{Syracuse Biomaterials Institute, Syracuse University, Syracuse, NY 13244\fnref{label2}}
  \address[label3]{{ICSMB, Department of Physics, University of Aberdeen, Aberdeen AB24 3UE, UK}\fnref{label3}}
  \address[label4]{{Martin Fisher School of Physics, Brandeis University, Waltham, MA 02453}\fnref{label4}}
  \address[label5]{{Instituto de Biocomputaci\'on y F\'isica de Sistemas Complejos (BIFI), 50009
Zaragoza, Spain.}\fnref{label5}}





\begin{abstract}

Minimal models of active Brownian colloids consisting of
self-propelled spherical particles with purely repulsive
interactions have recently been identified as excellent
quantitative testing grounds for  theories of active 
matter and have been the subject of extensive numerical and analytical
investigation. These systems do not exhibit aligned or
flocking states, but do have a rich phase diagram, forming
active gases, liquids and solids with novel mechanical 
properties. This article reviews recent advances in the 
understanding of such models,  including the description
of the active gas and its swim pressure, the motility-induced
phase separation and the high-density crystalline and
glassy behavior.
\end{abstract}


\end{frontmatter}


\section{Introduction}

Living entities, on scales from birds to individual cells, organize in complex patterns with collective behaviors that serve important biological functions. Examples range from the flocking of birds~\cite{ballerini:08} to the sorting and organization of cells in morphogenesis~\cite{friedl:09}. Work over the last ten years has shown that many aspects of this complex organization can be captured by physical models based on a minimal set of rules or interactions, leading to the emergence of the new field of active matter~\cite{marchetti:13}. This is defined as  a distinct category of non-equilibrium matter in which energy uptake, dissipation and movement take place at the level of discrete microscopic constituents. The active matter paradigm has additionally inspired the development of ingenious synthetic chemical and mechanical analogues, such as ``active'' colloids: micron-size spheres partly coated with a catalyst that promotes the decomposition of one of the components of the ambient fluid, resulting in self-propulsion of the colloidal particles~\cite{paxton:06,howse:07,palacci:10,poon:13}.  Collections of such active synthetic particles have been shown to spontaneously assemble in coherent  mesoscale structures with remarkable life-like properties~\cite{palacci:13,soto:15}.

Active systems exhibit rich emergent  behaviors, where a collection of many
interacting entities shows large-scale spatial or temporal organization in
states with novel macroscopic properties. For instance, a dense swarm of
bacteria can behave collectively as a living fluid with novel
rheology~\cite{lopez:15,marchetti:15}, self-organize in complex regular
patterns~\cite{budrene:95}, exhibit turbulent motion~\cite{dombrowski:04}, or
`freeze' into a solid-like biofilm ~\cite{hall-stoodley:04}. This type of behavior is of course well
known in inert inanimate matter that exhibits transitions between different
phases upon the tuning of an external parameter, such as temperature, or the
application of external forces that perturb the system at its boundaries (e.g.,
shear stresses) or globally (e.g., an electric or magnetic field).  It acquires, however,
a new unexplored richness in active systems that are tuned out of equilibrium
by energy generated internally by each unit. The active matter paradigm aims at
describing and classifying the behavior of this new class of non-equilibrium
systems. It does so by drawing on our understanding of familiar states of matter and of the
transitions between them  as controlled by interactions between atoms and
molecules. New states of matter arise when we put together many units that are
individually driven or motile. How are these new states formed? Are they
controlled solely by local interactions among the active particles or is
chemical signaling required to understand the emergence of these new states?
Can we classify and describe them and control the transitions between such
states as we know how to do with familiar inert matter? 
\begin{figure}[h]
\centering
\includegraphics[width=0.99\linewidth]{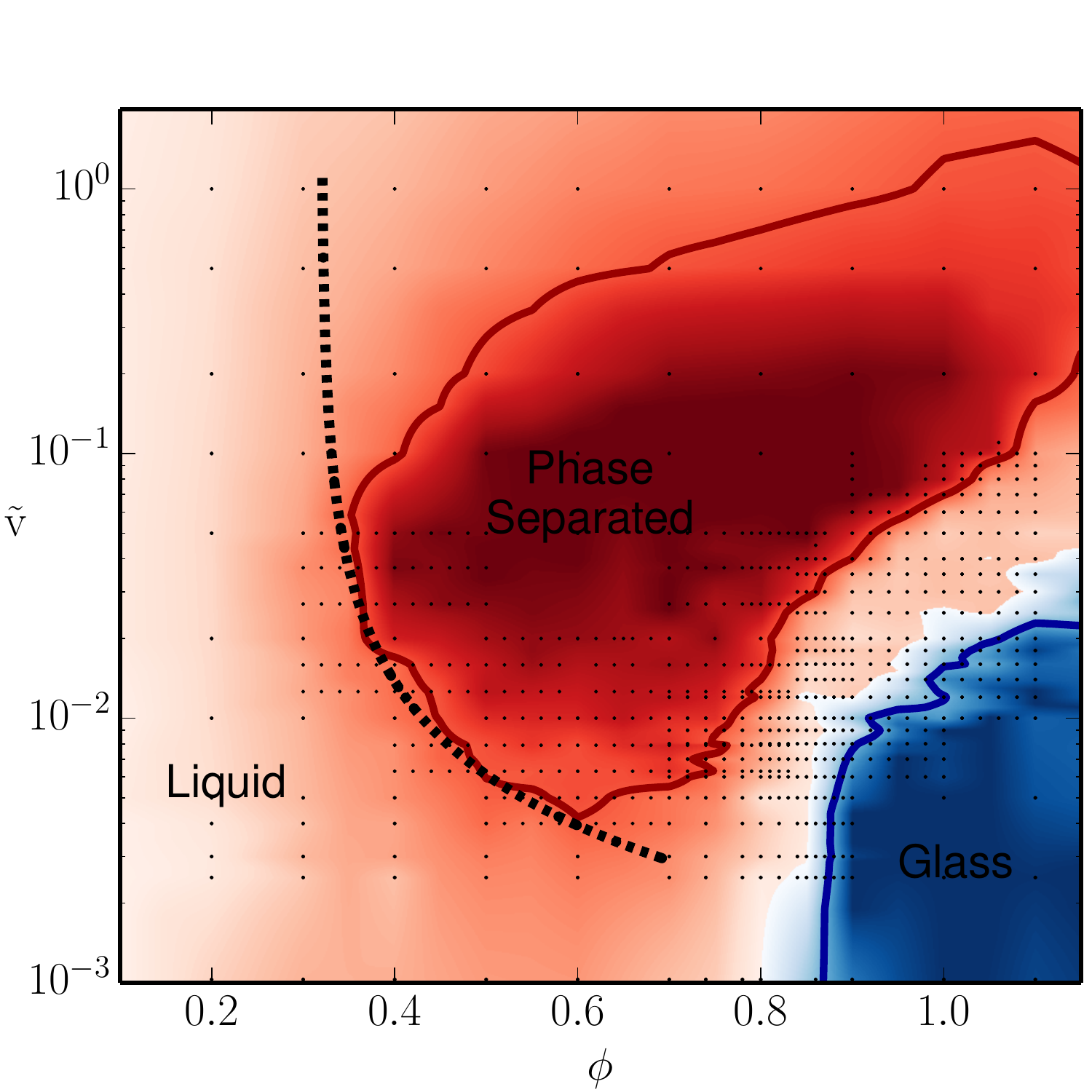}
\caption{
Numerical phase diagram of a polydisperse active suspension with soft repulsion
obtained by integrating Eqs~(\ref{eq:ri}--\ref{eq:thetai}) for $D_r/(\mu
k)=5\times10^{-4}$ and $D_t=0$, reproduced from Ref. ~\cite{fily:14} with
permission from the Royal Society of Chemistry.  The red region corresponds to
a phase separated system. The blue region corresponds to a glass as
characterized by the behavior of the MSD. The glass  would be replaced by a
crystalline state in a monodisperse suspension.  The dotted line is the
mean-field spinodal line given by $\rho_{c}^-(\Pe)$ from Eq.~(\ref{eq:phis})
for $D_t=0$.}
\label{fig:PD}
\end{figure}

{Recently, a number of ingenious synthetic systems have been engineered that show the emergent behavior of living active systems. 
These include autophoretic
colloids~\cite{howse:07,palacci:10,theurkauff:12},
rollers~\cite{bricard:13}, and droplets~\cite{thutupalli:11}. The simplest realization of such ``colloidal microswimmers'' is obtained by immersing spherical Janus colloids created by coating a hemisphere of a gold bead with platinum in a solution rich in hydrogen peroxide $(H_2O_2$)~\cite{paxton:06}. The difference in the consumption rate of $H_2O_2$ at the gold and platinum sides maintains an asymmetric concentration of solute on the two hemispheres, resulting in propulsion of the particles along their symmetry axis. In this and other catalytic colloidal swimmers
interactions and propulsion can be tuned in a controlled way, allowing systematic studies up to moderate densities (see Fig.~2(b). One remarkable phenomenon shown by these systems is spontaneous assembly in macroscopic clusters~\cite{theurkauff:12,palacci:13}. This phenomenon is distinct from equilibrium assembly that arises from attractive interactions between the  particles and has been shown to be driven by the nonequilibrium interplay of motility and crowding. The nonequilibrium pressure equation of state of active colloids has been probed experimentally via sedimentation measurements~\cite{ginot:15}, revealing a motility-induced effective adhesion that can strongly suppress the pressure at moderate density. A review of recent experimental findings can be found in Ref.~\cite{bialke:15} and is beyond the scope of the present article that focuses on minimal models of active colloids that have in many cases predicted and then qualitatively reproduced many of the experimental   observations.}

Colloids have played a key role in condensed matter physics as model systems
for atomic materials where pair interaction can be customized and equilibrium
phase transitions and glassy behavior can be investigated with optical
microscopy~\cite{lu-weitz:13}. Although many active entities, from bacteria to birds, are
elongated in shape and order in states with local or global liquid crystalline
order leading to collective flocking, the growing body of work on synthetic
active systems  has shown that
even spherical active particles can exhibit novel behaviors arising
from the irreversible dynamics of each constituent, providing an excellent
system for the quantitative testing of active matter theories.  This has led to
extensive theoretical and numerical studies of minimal models consisting of self-propelled {spheres} with purely repulsive interactions {known as active Brownian {particles (ABP)}}~\cite{cates:15,bialke:15}. These systems  do not exhibit
aligned or flocking states, but form active gases, liquids and solids, as summarized in the phase diagram shown in
Fig.~\ref{fig:PD}. The interplay between motility and steric effects is
responsible for intriguing new phenomena, including motility-induced phase
separation in the absence of any attractive
interactions~\cite{fily:12,cates:15}, Casimir-type forces~\cite{ray:14}, and
ratchet effects~\cite{wan:08}. {The complexity of behavior that arises in these minimal models is truly remarkable. Importantly, the ABP model has demonstrated that many aspects of the emergent behavior of active systems do not require biochemical signaling, but are captured by physical contact interactions. It has additionally provided} an excellent
playground for addressing fundamental questions about the non-equilibrium
statistical mechanics of active systems and whether equilibrium-like
notions, such as effective temperature or equations of state, may
be useful to describe them. 

In this article we review recent advances on the theoretical description of collections of  active Brownian particles (ABP) defined as spherical self-propelled particles with purely repulsive interactions  by organizing their behaviors in terms of the new active gases, liquids and solid phases formed by these systems. {The work described demonstrates that some key aspects of the emergent behavior of active systems, such as the tendency to spontaneously cluster in large compact structures and to accumulate at surfaces exerting organized forces on the environment, can be described in terms of the nonequilibrium interplay of motility and crowding, without invoking attractive interactions nor biochemical signaling. We believe that minimal models of the type described here will continue to provide important tools for advancing our understanding of the nonequilibrium statistical mechanics of active matter.}

The rest of this paper is organized as follows. In Section~\ref{model} the
minimal model of ABP is presented, emphasizing its
parameters, limiting cases, and critical values. Next, in
Section~\ref{idealgas}, we consider the properties of active gases  in the
context of recent work characterizing their mechanical properties and defining
a pressure equation of state. From the dilute ideal active gas limit, tuning
the rotational P\'eclet number and increasing density beyond a critical value results in
phase separation, where significant groups of active particles find their
self-propulsion velocity caged due to increased interactions, yielding an
active liquid coexisting with the active gas. This has been characterized by
equations of mean velocity and critical density, which are presented and
supported by  simulation. This motility-induced phase separation (MIPS) is
analyzed in Section~\ref{MIPS} using continuum equations. The high-density
limit, discussed in Section~\ref{crystals}, is one in which solids and glasses
of active colloids are formed, and this is distinguished from the 
glasses or crystalline states that results from high density collections of passive
colloids. Finally, a discussion and outlook is provided in
Section~\ref{discussion}, raising current questions regarding the interplay of
noise and damping in the current models of active systems.

\section{A minimal model of active colloids}
\label{model}

\begin{figure}[h]
\centering
\includegraphics[width=0.99\linewidth]{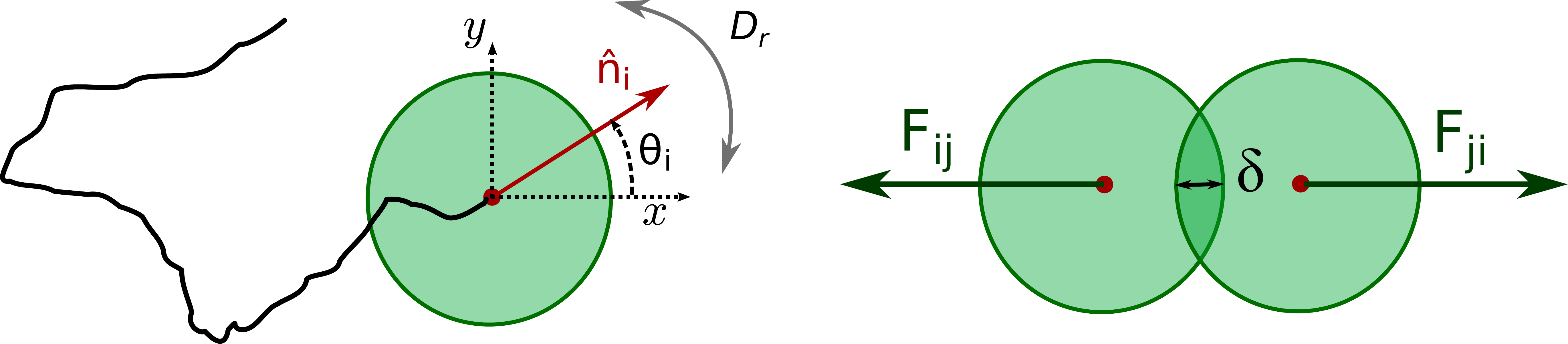}
\caption{{
Basic ingredients of the active Brownian particle model. 
Left: Each particle is self-propelled at speed $v_0$ along the direction  $\hat{\bf{n}}_i$ defined by the angle $\theta_i$ it makes with the $x$ axis. 
The angle $\theta_i$ is subject to angular noise determined by the diffusion rate $D_r$, corresponding to a persistence time $D_r^{-1}$. 
The jerky line that ends at the particle is a typical trajectory.
Right: The pairwise repulsive interaction is proportional to the overlap $\delta$.
}}
\label{fig:ABP_sketches}
\end{figure}

The rich behavior of active colloids has been studied using a minimal model of self-propelled particles (SPPs) that allows for both analytical and numerical progress. In this model  hydrodynamic interactions are neglected and the ambient fluid is assumed to only provide friction, rendering the dynamics overdamped. Each colloid is modeled as a spherical particle of radius $a_i$, with an orientation defined by the axis of self-propulsion. In the following we will discuss both monodisperse systems, where all disks have the same radius $a$, and polydisperse systems, where $a$ will denote the mean radius and the radii are uniformly distributed with $20\%$ polydispersity. While most of the work on this model known in the literature as ABP has been carried out in two dimensions ($2d$), and we will restrict ourselves  to this case here, many of the results described also hold in three dimensions~\cite{stenhammar:14}. Each particle is characterized by the position $\bm{r}_i$ of its center and its orientation $\bm{e}_i=\left(\cos\theta_i,\sin\theta_i\right)$ which in $2d$ corresponds to a single angle $\theta_i$. The dynamics is then described by coupled Langevin equations {(see figure~\ref{fig:ABP_sketches})}
\begin{eqnarray}
\label{eq:ri}
&&\partial_t\bm r_i=\vel_0\bm{e}_i+\mu\sum\limits_{j} \bm f_{ij}+\bm\eta_i(t)\;,\\
\label{eq:thetai}
&&\partial_t\theta_i=\eta^r_i(t)\;,
\end{eqnarray}
where $\vel_0$ is the active (self-propulsion) speed and $\mu$ the mobility.  The particles interact via short-range radial repulsive forces $\bm f_{ij}=-\frac{\partial U(|\bm{r}_{ij}|)}{\partial\bm{r}_i}$, where $U(r)$ is a pair interaction that depends on the separation $\bm{r}_{ij}=\bm{r}_{i}-\bm{r}_{j}$ of the centers of the two particles.  Both soft and hard repulsive forces have been used in the literature, and the qualitative behavior of the system is not affected by the details of the pair repulsion, provided it is pairwise, radial, and decoupled from the angular dynamics. Here we assume soft repulsive forces $\bm{f}_{ij}=f_{ij}\bm{\hat{r}}_{ij}$, with 
$\bm{\hat{r}}_{ij}=(\bm{r}_i-\bm{r}_j)/r_{ij}$, $r_{ij}=|\bm{r}_i-\bm{r}_j|$, and ${ 
f}_{ij}=k(a_i+a_j-r_{ij})$ if $r_{ij}<a_i+a_j$ and ${f}_{ij}=0$ otherwise. We consider $N$ disks in an area $A=L^2$, with $\rho=N/A$ the number density and $\phi=\sum_i\pi a_i^2/A$ the  packing fraction.

The random force in Eq.~(\ref{eq:ri}) describes thermal noise, with zero mean and correlations 
$\langle\eta_{i\alpha}(t)\eta_{j\beta}(t')\rangle=2D_t\delta_{\alpha\beta}\delta_{ij}\delta(t-t')$, where $D_t= k_\text{B}T\mu$ is the translational diffusion coefficient and $k_\text{B}T$ the thermal energy. The direction of self propulsion fluctuates due to non-thermal processes. This is described by the random torque  $\eta^r_i$ which is also chosen with zero mean and correlations $\langle\eta^r_i(t)\eta^r_j(t')\rangle=2D_\text{r}\delta_{ij}\delta(t-t')$, with $D_\text{r}$  the rotational diffusion rate, which is  treated as an independent parameter because in many realizations, including bacterial suspensions~\cite{berg:04} and  active colloids~\cite{palacci:13}, the rotational noise is athermal.  

It is instructive to note that Eqs.~(\ref{eq:ri}) and (\ref{eq:thetai}) can be combined into a single equation of the form~\cite{fily:12}
\begin{eqnarray}
\label{eq:ri-one}
\partial_t\bm r_i=\bm\xi_i+\mu\sum\limits_{j} \bm f_{ij}+\bm\eta_i(t)\;,
\end{eqnarray}
where $\bm\xi_i(t)=\vel_0\bm{e}_i(t)$ is a non-Markovian noise with zero mean and correlations
\begin{equation}
\label{eq:xi}
\langle\xi_{i\alpha}(t)\xi_{j\beta}(t')\rangle=D_\text{a}\delta_{\alpha\beta}\delta_{ij}\frac{\text{e}^{-|t-t'|/\tau_\text{r}}}{\tau_\text{r}}
\end{equation}
with $D_\text{a}=\vel_0^2\tau_\text{r}/2$ the diffusion constant of the
persistent motion. In other words the active forcing on the right-hand side of
Eq.~(\ref{eq:ri}) is equivalent to non-Markovian noise with a persistence time
$\tau_\text{r}=1/D_\text{r}$. This key property of ABP arises because in this
minimal model of active colloids the single-particle angular dynamics is
decoupled from interactions and from the angular dynamics of other particles. {As a result, ABPs do not exert torques on each other nor on the walls of a container. This simplification has allowed substantial theoretical progress, but has important limitations discussed in Section 6.}
Note that Eqs.\ref{eq:ri} and  \ref{eq:xi} are distinct from Gaussian colored
noise due to the bounded speed distribution here. 
In the limit $\tau_\text{r}\!\rightarrow 0$, for fixed $D_\text{a}$, the exponential decay can be approximated by a delta function, 
$\left(\text{e}^{-|t-t'|/\tau_\text{r}}\right)/\tau_\text{r}\!\rightarrow\delta(t-t')$. In this limit, a single active Brownian particle behaves like a thermal Brownian particle with an effective temperature $k_\text{B}T_\text{eff}=\frac{\vel_0^2}{2\mu D_\text{r}}$~\cite{fily:12}. 

In interacting ABP the thermal limit is only realized, however, if the
orientational correlation time $\tau_\text{r}$  is much smaller than all time
scales present in the system.  In particular, $\tau_\text{r}$ must be small
compared to the mean free time between collisions,  $\tau_\text{f}\simeq(2a
\vel_0\rho)^{-1}$,  and to the interaction time $\tau_k=(\mu k)^{-1}$.  The
dimensionless ratio $\zeta=\tau_\text{r}/\tau_\text{f}$ is especially relevant
for phase separation, whose onset has been interpreted in terms of a critical
value of $\zeta$~\cite{redner:13}.  A recent study has also pointed out the
crucial role played by $\zeta$ in active suspensions, where hydrodynamic
interactions are important~\cite{matas-navarro:14}.

We will see below that ABP can be tuned through active gas, liquid and solid
phases by tuning activity, as measured by a dimensionless rotational P\'eclet
number $\Pe=\vel_0\tau_\text{r}/a$ and the packing fraction $\phi$~\footnote{We
stress that Refs.~\cite{takatori:14,takatori:14b} characterize activity in
terms of a translational P\'eclet number defined in terms of the diffusivity
$D_\text{a}$, hence proportional to $1/\vel_0$. Here we prefer to use the
rotational P\'eclet number, which is commonly used in the active matter
literature and is a direct measure of persistence.}.

\section{Active gases}
\label{idealgas}
\begin{figure*}[t!]
\centering
\begin{minipage}[t]{.45\linewidth}
\centering
{\Large (a)}

\includegraphics[height=5.33cm]{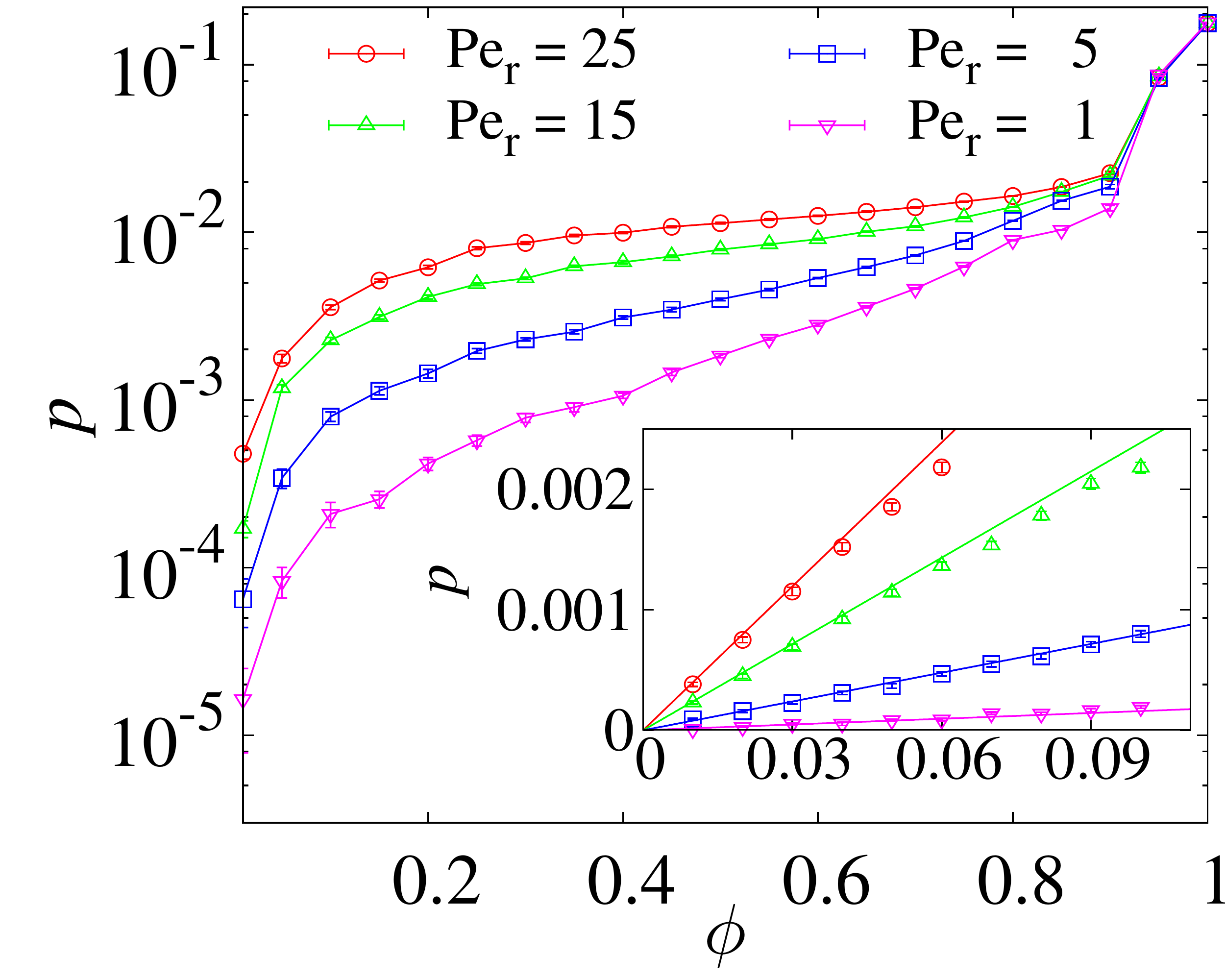} 
\end{minipage}
\begin{minipage}[t]{.45\linewidth}
\centering
{\Large (b)}

\includegraphics[height=5cm]{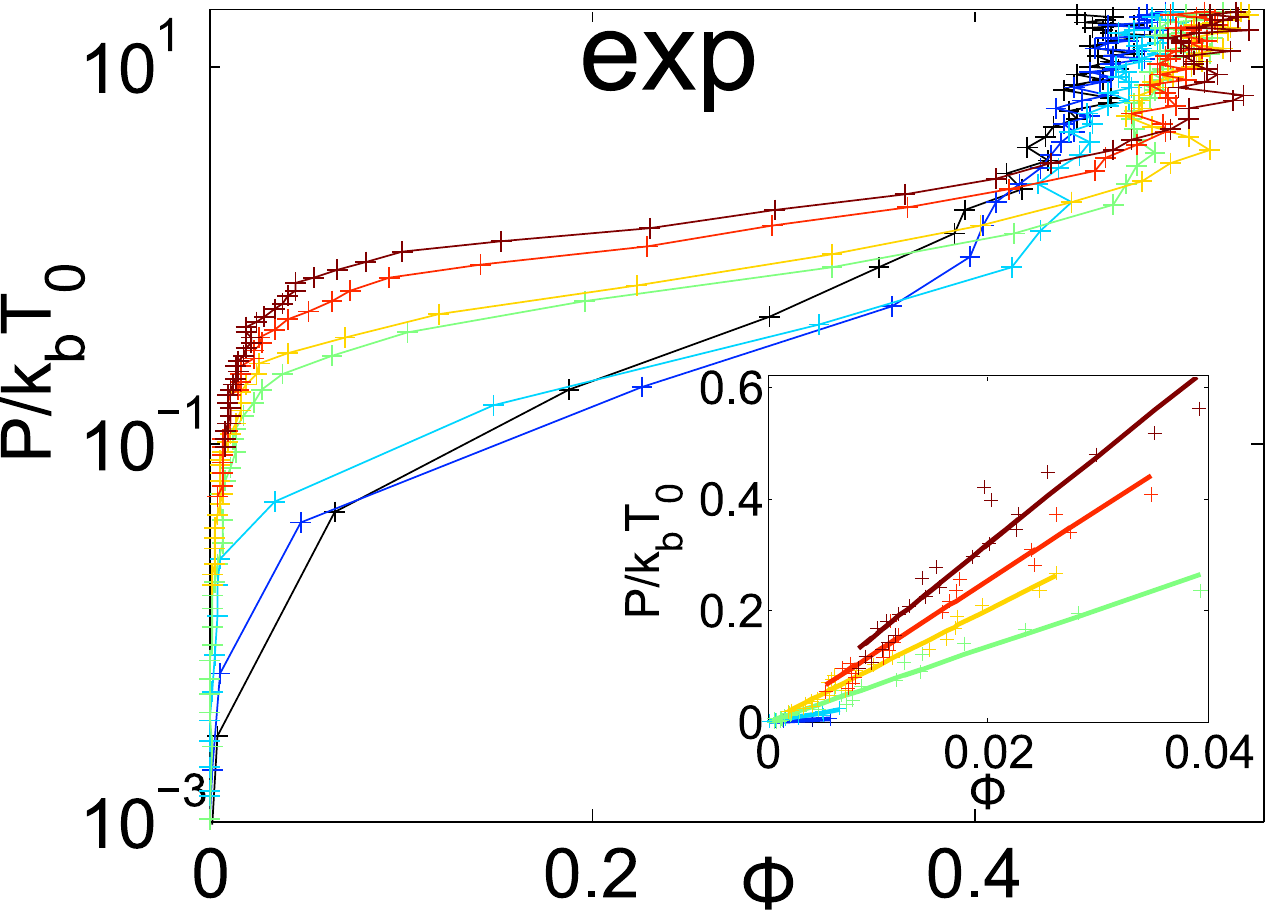}
\end{minipage}

\begin{minipage}[t]{.45\linewidth}
\centering
{\Large (c)}

\includegraphics[height=4.8cm]{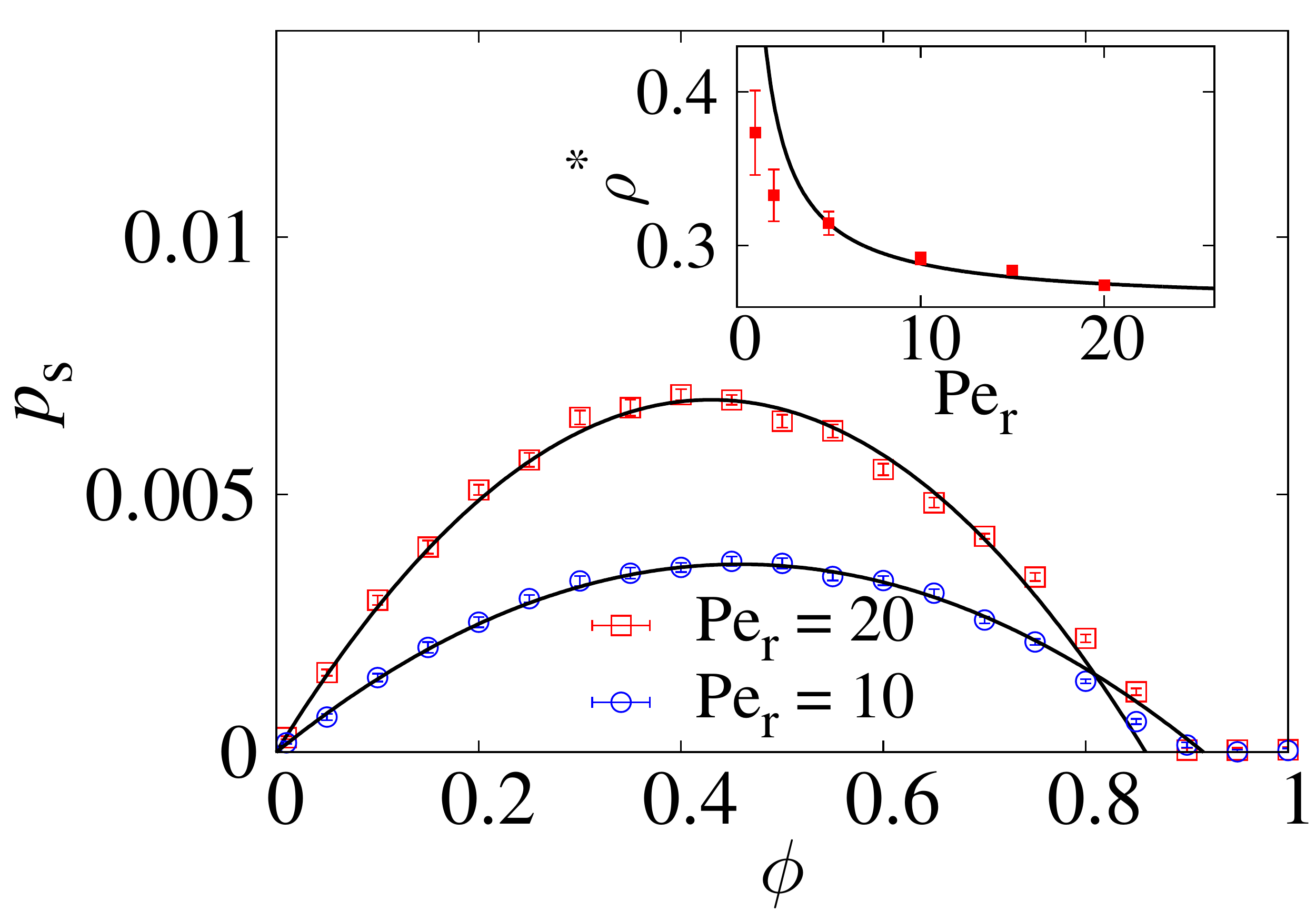} 
\end{minipage}
\begin{minipage}[t]{.45\linewidth}
\centering
{\Large (d)}

\includegraphics[height=4.8cm]{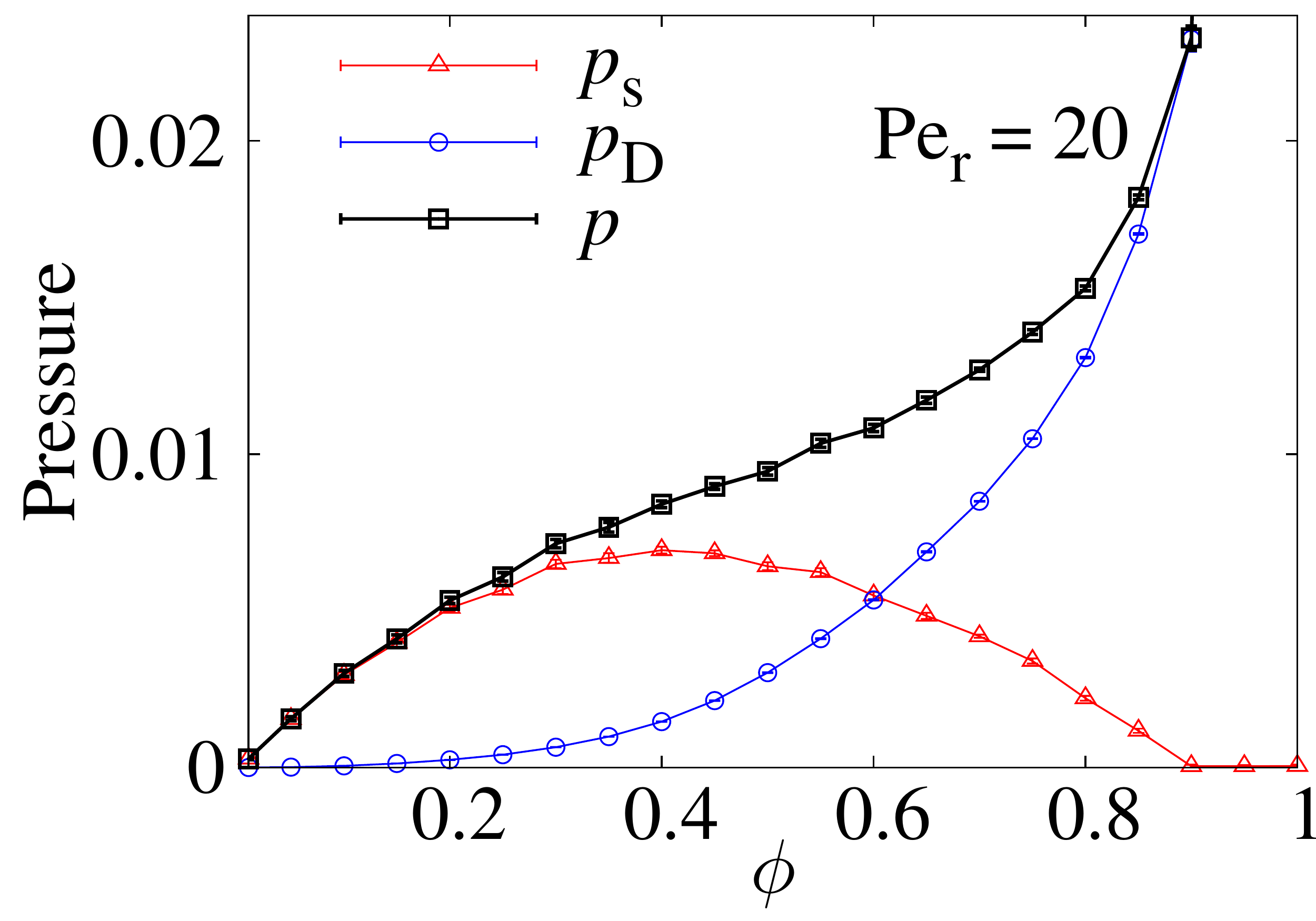}
\end{minipage}
\caption{
Pressure as a function of packing fraction $\phi$ from (a) our simulations of
self-propelled particles with soft repulsive interactions for increasing values
of $\Pe=\ell_\text{p}/a$,  obtained by decreasing $D_\text{r}$ at fixed $v_0$ ,
and (b) sedimentation experiments of active Janus colloids by Ginot \emph{et
al.}~\cite{ginot:15} (reproduced with permission form Ref.~\cite{ginot:15}).
{These experiments use gold microsphere half-coated with platinum and immersed in a bath of $H_2O_2$.  The suspension is at an ambient temperature $T_0=300K$  and the various curves corresponding to
increasing concentration of $H_2O_2$ resulting in increasing self-propulsion
speed (from bottom to top). 
The pressure is extracted form measured density profiles in colloids sedimenting under gravity in a slightly tilted geometry that allows to control and reduce the strength of gravity.  }
 The lines in frame (a) are a guide to the eye. In
both (a) and (b) the insets show the pressure for a dilute active gas. {In both insets the straight lines are a fit to Eq.~(\ref{eq:Pideal}) (augmented by the thermal ideal gas pressure in Fig. 2(b)) with no adjustable parameters.}    Frame (c) shows the swim pressure vs $\phi$ for $\Pe=10,20$. The
solid lines are fits to Eq.~(\ref{eq:ps2}) using
$\vel(\rho)=\vel_0(1-\rho/\rho_*)$ (for each \Pe\ $\rho^*$ is
the only fit parameter).  The inset shows $\rho_*(\Pe)$ and a fitfbasic
to~(\ref{eq:rhostar}).  Frame (d) displays the various contributions to the
pressure for $\Pe=20$.   In all the simulations, $\tau_k =(\mu k)^{-1}=1$ is
chosen as the time unit, $a=1$, $v_0=0.01$ and $D_\text{r}$ is varied to obtain
the desired \Pe.  We follow the system for a time of $10^6 \tau_k$, or
equivalently at least $2\times 10^3 \tau_r$, more than enough to reach the
steady state, and average over several dozen runs to obtain our error
estimates.
\label{fig:pressure}}
\end{figure*}
It is useful to first examine the behavior of an ideal gas of SPPs. In the
absence of interactions, each particle performs a persistent random walk. The
dynamics is characterized by the persistence time $\tau_\text{r}$ of the path
and the associated persistence length $\ell_\text{p}=\vel_0\tau_\text{r}$. The rotational P\'eclet number also represents the ratio of the persistence length to the size of the particles,
$\Pe=\ell_\text{p}/a$ and provides a
measure of persistence or activity. The mean square displacement (MSD)
of a single SPP can be calculated exactly and it is given by
\begin{equation}
\label{eq:MSD}
\langle[\Delta\bm{r}(t)]^2\rangle=4D_tt+2\vel_0^2\tau_\text{r}\left[t-\tau_\text{r}\left(1-\text{e}^{-t/\tau_\text{r}}\right)\right]\;,
\end{equation}
with $\Delta\bm{r}(t)=\bm{r}(t)-\bm{r}(0)$.
The dynamics is ballistic for $t\ll\tau_\text{r}$, with
$\langle[\Delta\bm{r}(t)]^2\rangle\sim \vel_0^2t^2$, and diffusive for
$t\gg\tau_\text{r}$, with $\langle[\Delta\bm{r}(t)]^2\rangle\sim 4[D_t+D_a]t$, and
$D_a=\vel_0^2/(2D_\text{r})=\ell_\text{p}^2/(2\tau_\text{r})$.  The same expression is obtained for
run-and-tumble bacteria, with the replacement $D_\text{r}\rightarrow\alpha$ and
$\alpha$ the tumble rate, indicating that the coarse-grained dynamics is
insensitive to whether changes in direction of self-propulsion are continuous
or discrete. A detailed comparison of rotational and run-and-tumble dynamics
can be found in Ref. \cite{solon:15c}. For both bacteria and active
colloids the thermal diffusion coefficient $D_t$ is typically two orders of
magnitudes smaller than the active diffusion $D_a$. This justifies neglecting
thermal noise in Eq. (\ref{eq:ri}). 

Recently there has been substantial theoretical effort towards characterizing
the mechanical properties of an active gas and understanding whether this
non-equilibrium system can be described by a pressure equation of
state~\cite{mallory:14,yang:14,takatori:14,solon:15,solon:15b}.  In
equilibrium, the pressure $p$ of a fluid can be defined in three equivalent
ways:
(i) as the mechanical force per unit area on the walls of the container, (ii)
from thermodynamics as the derivative of a free energy, and (iii)  as the trace
of the hydrodynamic stress tensor of the fluid, which in turn represents the
momentum flux in the system. In equilibrium all three definitions give the same
expression and the pressure is  a state function.  Recent work by us and others
has shown that the pressure is also a state function for an active fluid of
spherical SPPs with purely repulsive
interactions~\cite{yang:14,takatori:14,solon:15}, but becomes dependent on
the nature of the walls as soon as the active particles exert torques on each
other or on the boundaries, as would for instance be the case for non-spherical
SPPs~\cite{solon:15b}. 

{Starting with the familiar virial expression~\cite{hansen:06}, the contribution from interparticle interactions to the pressure $p_D$ of a passive fluid with pairwise forces $\bm{f}_{ij}$ can be written as the trace of the stress tensor~\cite{irving:50} (see, e.g., Section~8.4 of~\cite{hansen:06}),}
\begin{equation}
\label{eq:pD}
p_D=\biggl\langle\frac{1}{2A}\sum_{i<j}\bm{f}_{ij}\cdot\bm{r}_{ij}\biggr\rangle\;,
\end{equation}
where $A$ is the area of the system and  the brackets denote an average over the noise. {This expression, often referred to as Irving-Kirkwood formula~\cite{irving:50}, holds generally under the assumption of pairwise interactions } and highlights the meaning of pressure as describing moment flux across a unit line (or a unit plane in three dimensions).
 For passive gases in thermal equilibrium to this direct contribution from interaction  one must add the momentum carried by free streaming particles, which gives the ideal gas pressure $p_0=\rho k_\text{B}T$, with $\rho=N/A$ the number density.   In a fluid of SPPs there is an additional contribution to the pressure  that describes the flux of self-propulsive force across a unit line. This contribution, unique  to active systems, was first identified in Refs.~\cite{mallory:14,yang:14,takatori:14} and was dubbed `active' or `swim' pressure. We will use the latter term here that has now become accepted in the literature. In Ref.~\cite{yang:14} it was shown that the swim pressure can be written in a virial-type form as
\begin{equation}
\label{eq:ps}
p_s=\biggl\langle\frac{1}{2A}\sum_i \bm{F}^a_i\cdot\bm{r}_i\biggr\rangle\;,
\end{equation}
where $\bm{F}^a_i=\vel_0\bm{e}_i/\mu$ is the propulsive force on each particle.
The total pressure of an active gas  is then $p=p_0+p_D+{p_s}$. In the remainder
of this section we will consider active systems at $T=0$ and ignore the thermal
ideal gas contribution. It was shown in Ref.~\cite{yang:14}  that for repulsive
self-propelled disks the pressure calculated as the sum of Eqs.~(\ref{eq:pD})
and (\ref{eq:ps}) is indeed identical to the force per unit area on the walls
of the container, demonstrating that in this system the pressure is a state
function.   When the active units exchange torques among themselves and/or with
the walls of the container, as is the case for non-spherical particles or in
the presence of orientation-dependent interactions, the force on a wall depends
on the properties of the wall and the definition of a mechanical state function
that plays the role of pressure does not seem possible~\cite{solon:15b}. 

In the absence of interactions $p_D=0$ and  one can  readily calculate the
pressure $p_s^0$ of an active ideal gas using Eq.~(\ref{eq:ps}) and assuming
that the initial position of the particle is not correlated with the noise,
with the result $p_s^0(t)=\frac{\rho \vel_0^2}{2\mu
D_\text{r}}\left(1-\text{e}^{-D_\text{r}t}\right)$. In the limit
of large systems, one can let $t\rightarrow\infty$ in this expression to obtain
the pressure of an ideal active gas in the thermodynamic limit,
\begin{equation}
\label{eq:Pideal}
p_s^0=\rho \frac{\vel_0^2}{2\mu D_\text{r}}=\rho k_\text{B}T_\text{eff}\;.
\end{equation} 
In other words the swim pressure of an ideal active gas is simply the pressure
of an ideal gas at the temperature $T_\text{eff}$. Although
Eq.~(\ref{eq:Pideal}) replaces the kinetic contribution to the pressure that
one would have in a thermal passive gas,  we will see below that, unlike the
ideal gas pressure of thermal systems, this contribution is strongly
renormalized by interactions.  Additionally, active systems are much more
sensitive to finite size effects and care is required when taking the
thermodynamic limit.  If the persistence length $\ell_\text{p}$ is comparable
to or exceeds the linear size $L$ of the container, then particles travel
ballistically and one can estimate the pressure   as $p_a^0(t\sim
L/\vel_0)=\frac{\rho \vel_0^2}{2\mu
D_\text{r}}\left(1-\text{e}^{-L/\ell_\text{p}}\right)\simeq \frac{\rho
\vel_0aL}{2\mu}$ where the last approximate equality holds when
$\ell_\text{p}>> L$. For high activity, the active gas behaves like a highly
diluted Knudsen gas and the pressure depends on the size of the container.  It
has also been shown recently that boundary curvature can strongly affect the
density and force distribution in ABP~\cite{fily:14b}.  Care must therefore be
taken in numerical simulations as the mapping of a dilute active gas of purely
repulsive spherical colloids onto a thermal gas at an effective temperature
$T_\text{eff}=\vel_0^2/(2\mu D_\text{r})$ only holds when $\ell_\text{p}\ll L$.
Strong finite-size effects are observed when $\ell_\text{p}$ is a fraction of
$L$. In this case active gases behave quite differently from thermal ones: they
accumulate at the walls of the container \cite{yang:14} or around fixed
impurities, exert Casimir-type forces~\cite{ray:14}, and spontaneously sort
when different in size or activity~\cite{yang:14}. P\'eclet numbers in excess
of $100$, corresponding to persistence lengths of order $100$ particle sizes,
are easily reached in experiments in active colloids~\cite{buttinoni:13} and
finite size effects should be observable in microfluidic devices. 

At higher density, interactions become important. It is instructive
to note that the kinetic contribution to the pressure of thermal Brownian
particles can also be calculated through a  virial-type expression akin to the
one defining the swim pressure of ABP, given by
$p_0=\lim_{t\rightarrow\infty}\frac{1}{2A}\langle\sum_i\mathbf{\eta}_i\cdot\mathbf{r}_i\rangle$,
with $\mathbf{\eta}_i$ the thermal noise. Assuming, as customary, that the
initial position of the particles is not correlated with the noise, and making
use of the fact that the noise is uncorrelated in time, this expression gives
exactly $p_0=\rho k_BT$ at all densities $\rho$. In contrast, the stochastic
propulsive force entering the definition of the swim pressure,
Eq.~(\ref{eq:ps}), is correlated in time (see Eq.~(\ref{eq:xi})), leading to
the dependence of the swim pressure on interparticle forces.  The total
pressure of a monodisperse system of repulsive SPPs is shown in
Fig.~\ref{fig:pressure}(a) as a function of packing fraction for several values
of the persistence length. At low density the system behaves like a gas of
`hot' repulsive colloids with effective temperature $T_\text{eff}$, as shown in
the inset. At intermediate density the swim pressure is strongly suppressed due
to collisional slowing down (Fig.~\ref{fig:pressure}(c)) and the total pressure
remains almost constant over a range of packing fractions. This  is the region
where the system spontaneously phase separates, as discussed below, and a fluid
of repulsive SPPs behaves qualitatively like a passive fluid of attractive
particles. At high density the swim pressure is negligible and repulsive
interactions dominate, yielding the sharp increase of the pressure associated
with the transition to a glassy or solid state, as in passive systems.  The
pressure of active Janus colloids was recently measured in sedimentation
experiments~\cite{ginot:15}, confirming the prediction of the simulations (see
Fig. \ref{fig:pressure}(b)).  These experiments also confirmed that dilute
active gases behave like `hot' passive colloids with an effective temperature
enhanced by activity and that this mapping holds  even in the presence of
interactions up to moderate densities, but below the region of spontaneous
phase separation. 

It was shown in Ref.~\cite{solon:15} that in the thermodynamic limit  the swim pressure given in Eq.~(\ref{eq:ps}) can  be recast in an intuitive form, given by
\begin{equation}
\label{eq:ps2}
p_s=\rho\frac{\vel_0\vel(\rho)}{2\mu D_\text{r}}\;,
\end{equation}
where $\vel(\rho)$ is the mean velocity of the particles along the direction of self-propulsion, given by 
$\vel(\rho)=\vel_0+\mu\langle\bm{e}_i\cdot\sum_{j\not=i}\bm{f}_{ij}\rangle$. We note that this expression only holds if $\ell_\text{p}\ll L$ and in fact reduces to the bulk pressure of an active ideal gas, $p_s^0=\rho \vel_0^2/(2\mu D_\text{r})$, in the absence of interactions.  The suppression of the swim pressure due to collisional slowing down can then be modeled phenomenologically at moderate densities as a linear decay of the propulsion speed, with $\vel(\rho)=\vel_0(1-\rho/\rho_*)$ for $\rho<\rho_*$ and $\vel(\rho)=0$ for $\rho>\rho_*$. In general $\rho_*$ will depend on $\Pe$.
This linear decay of $\vel(\rho)$ with $\rho$ has been seen in simulations \cite{fily:12} and justified by relating the linear decay rate to the pair correlation function \cite{bialke:13} and by estimating it via a kinetic argument  \cite{stenhammar:13,fily:14} as arising from the fact that at finite density particles can be stalled for the duration $\tau_\text{c}$ of collisions with other particles. At low density, where the mean free time between collisions $\tau_f=(v_0 2a\rho)^{-1}$ exceeds $\tau_\text{c}$, the mean speed can then be written as 
$\vel(\rho)= \vel_0(1-\tau_\text{c}/\tau_f)$. The collision time $\tau_\text{c}$ was treated as a constant fitting parameter in Ref.~\cite{stenhammar:13}, but will in general depend on P\'eclet number. 
It can be estimated as controlled by two delay mechanisms:   the time $a/\vel_0$ it takes two interacting particles to move around each other and the reorientation time $D_\text{r}^{-1}$. Since the collision time $\tau_\text{c}$ will be controlled by the faster of these two processes, we add the two rates to obtain $\tau_\text{c}^{-1}\sim \vel_0/a+D_\text{r}$. This gives $\vel(\rho)=\vel_0\left[1-\rho/\rho_*(\Pe)\right]$, with
\begin{equation}
 \rho_*(\Pe)=\frac{c}{a^2}\left(1+\frac{1}{\Pe}\right)\;,
 \label{eq:rhostar}
 \end{equation}
where $c$ is a number of order unity. In our simulations (see
Fig.~\ref{fig:pressure}), we find that this expression works well for
$\Pe>2$,  with $c\approx0.26$. The swim pressure for moderately dense gases
can then be written as $p_s=\rho\frac{\vel_0^2}{2\mu
D_\text{r}}\left[1-\frac{c\phi}{\pi}\left(1+1/\Pe\right)^{-1}\right]$. At low
density, the direct part of the pressure can be estimated as
$p_D=\rho(\vel_0/\mu)\rho a^3$, where $v_0/\mu$ is the force scale and $\rho a^3$
is the typical displacement of a particle between collisions
\cite{takatori:14b}.  This expression holds as long as the particles do not
overlap.

\section{Active fluids and motility-induced phase separation}
\label{MIPS}
The most striking phenomenon exhibited by our minimal model of active colloids
is motility-induced phase separation (MIPS), where an active fluid of particles
with purely repulsive interactions spontaneously phase separates into a dense
liquid surrounded by an active gas~\cite{fily:12}. This remarkable phenomenon
arises from the collisional slowing down of the active particles' speed
$\vel(\rho)$ that yields the suppression of the swim pressure shown in Fig.
\ref{fig:pressure}(c), combined with the fact that the lack of detailed balance allows self-propelled particles to accumulate in the regions where $\vel(\rho)$) is small~\cite{schnitzer:93,tailleur:08}.  While experiments
often see strong clustering, but not complete phase separation
~\cite{theurkauff:12,palacci:13}, MIPS has now been seen in numerical
simulations  for a variety of repulsive interaction, ranging from soft harmonic
potential to WCA and hard spheres, in both two and three
dimensions~\cite{bialke:15,wysocki:14,stenhammar:14}.

The phase behavior of the system, as obtained from simulations of the
polydisperse limit of the model described in Section \ref{model}, is summarized
in Fig.~\ref{fig:PD}. Three regimes can be identified: a homogeneous fluid
phase, a solid or frozen state at high packing fraction and relatively low
activity, and a regime where the system shows macroscopic phase separation into
two bulk phases. The phase separated regime  exists in an intermediate range of
packing fraction and activity, corresponding to the region where the pressure
remains approximately constant with increasing packing fraction. For soft repulsion, it changes
continuously from a high density liquid cluster surrounded by a gas at lower
packing fractions, to a ``hole'' of gas phase inside a densely packed liquid at
higher packing fractions~\cite{fily:14}. The existence of a homogeneous liquid
phase intermediate between the solid and the phase separated state seems to be
a generic feature, but is not fully understood. In contrast, the suppression of
phase separation at high self-propulsion speed is specific to the soft nature
of the repulsive potential used in our simulations, where for $\vel_0/a \mu k>1$
particles can just pass through each other.

The threshold $\rho_\text{c}(\Pe)$ for phase separation has been estimated in various
ways in the literature.  A naive estimate is obtained by assuming that phase
separation occurs when $\tau_f<\tau_\text{r}$, corresponding to the condition that a
particle experiences many collisions before changing direction, i.e., it gets
trapped by other particles, with the result $\rho_\text{c}(\Pe)\sim (a^2 \Pe)^{-1}$. The
same estimate was obtained by Redner \emph{et al.} \cite{redner:13} via a
kinetic argument that equates the flux $\rho_{gas}\vel_0$  of particles that
enter the cluster to the flux out of the cluster, $\kappa D_\text{r}/a$, with $\kappa$
a fitting parameter.  The dependence on $D_\text{r}$ arises because particles must
turn around to point outward before they can leave the cluster.  By equating
these fluxes one can estimate the density of the gas as $\rho_{gas}=\kappa
D_\text{r}/(a \vel_0)$ and calculate the fraction $f_\text{c}$ of particles in the cluster as
a function of $\rho$ and $\Pe$. The condition $f_\text{c}=0$ gives a criterion for the
phase separation, corresponding to $\rho_\text{c}(\Pe)=\kappa/(a^2 \Pe)$. These naive
estimates predict that for arbitrarily large $\Pe$ phase separation will occur
no matter how low the density. This  behavior is not, however, born out by
simulations that find a minimum threshold of density below which there is no
phase separation.

\begin{figure}[h!] \centering
\includegraphics[width=0.99\linewidth]{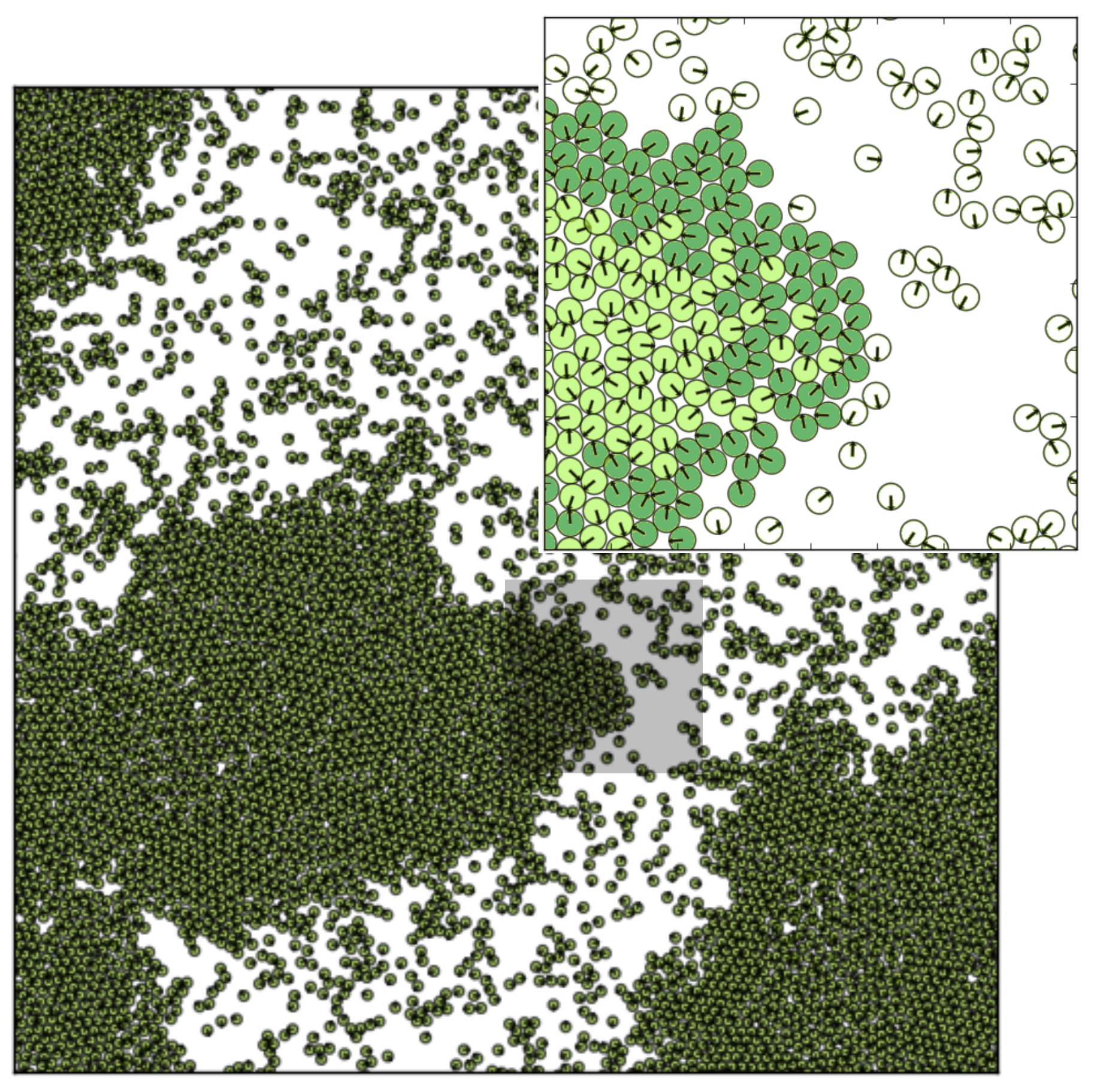}
\caption{Snapshot of an ABP simulation with inset zooming in on the
shaded region of the background image. The inset shows  particles in the dense liquid phase tagged with
green (dark green for inward-pointing boundary particles, light green for particles in the bulk of the dense liquid), while gas particles are tagged with white.
The self-propulsion vector is shown originating from the center of each disk. The full system shown in the background has periodic boundaries, $\phi=0.5$ and $L=200$.
The parameters are $\mu=k=1$, so $\tau_k=(\mu k)^{-1}$ is the time
unit, $v_0=0.01$ and $a=1$. The rotational diffusion has been set to give a
P\'eclet number of $\Pe=50$.  } \label{fig:cluster} \end{figure}

Another  approach is to use continuum equations for an active fluid where
motility suppression is incorporated through the density-dependent propulsion
speed $\vel(\rho)$ that controls the swim
pressure~\cite{tailleur:08,fily:12,cates:13,stenhammar:13,bialke:13}, and
then employ  linear stability analysis to locate the spinodal.  The dynamics is
governed by equations for the density $\rho({\bm r},t)$ and the  polarization
density ${\bm p}({\bm r},t)$ that describes the local orientation of the
particles' axis of self-propulsion,  given by~\cite{fily:12} 
\begin{eqnarray} \label{hydro} \label{rho}
&&\partial_t\rho=-\bm\nabla\cdot\left[\vel(\rho){\bm
p}-D_t\bm\nabla\rho\right]\;,\\ \label{p} &&\partial_t{\bm p}=-D_\text{r}{\bm
p}-\bm\nabla\pi(\rho)+K\nabla^2{\bm p}\;, \end{eqnarray}
where $\pi(\rho)=\vel(\rho)\rho/2=p(\rho)\mu D_\text{r}/\vel_0$, with $p(\rho)$ the
pressure.  For times $t\gg D_\text{r}^{-1}$, one can neglect the time derivative of
the polarization in Eq.~(\ref{p}) relative to the damping term and eliminate
$\bm{p}$ to obtain a  nonlinear diffusion equation for the density, given by
\begin{equation} \label{diff} \partial_t\rho=\bm\nabla\cdot\left[{\cal
D}(\rho)\bm\nabla\rho\right]\;, \end{equation}
with effective diffusivity 
\begin{equation} \label{Dcal} {\cal
D}(\rho)=D_t+\frac{\vel^2(\rho)}{2D_\text{r}}\left(1+\frac{\text{d}\ln
\vel}{\text{d}\ln\rho}\right)\;.  \end{equation}
The linear stability of a homogeneous state of constant density
$\overline{\rho}$ can be analyzed by examining the dynamics of density
fluctuations, $\delta \rho=\rho-\overline{\rho}$. 
The decay of fluctuations is controlled by a diffusive mode with diffusion
constant ${\cal D}(\overline{\rho})$ that can become negative, signaling the
instability of the uniform state.  The onset of the instability corresponds to
${\cal D}(\overline{\rho})=0$, which determines the spinodal line. When $D_t=0$
the threshold for phase separation in
$\rho_\text{c}(\Pe)=\rho_*(\Pe)/2=(2a)^{-2}\left(1+\Pe^{-1}\right)$. This expression does
yield a minimum density below which there is no phase separation, as seen in
simulations ~\cite{fily:14}, but it fails to account for the observed lower
bound for the value of $\Pe$ required for phase separation. A lower bound is
obtained when $D_t\not=0$. In this case  one finds
 \begin{equation}
\label{eq:phis}
\rho_{c}^\pm(\Pe)=\frac{\rho_*(\Pe)}{4}\left(3\pm\sqrt{1-\left(\frac{\Pe^\text{c}}{\Pe}
\right)^2}\right)\;,
 \end{equation} 
where $\Pe^\text{c}=4\sqrt{D_t/(a^2D_\text{r})}$
represents the  minimum value of $\Pe$ required for phase separation. Thermal
diffusion suppresses phase separation and drives this critical $\Pe$ to larger
values. The condition $\Pe>\Pe^\text{c}$ for phase separation can also be written as
$D_\text{r}<\vel^2/a^2D_t$, indicating that the observed spinodal decomposition
requires a sufficiently small $D_\text{r}$ or long persistence time. The expression
given in Eq.~(\ref{eq:phis}) with the choice of minus sign corresponds to the
dotted line in Fig.~\ref{fig:PD}. It is, however, unclear whether thermal
diffusion is really the mechanism responsible for the lower bound on $\Pe$,
given the simulations leading to Fig.~\ref{fig:PD} were carried out with
$D_t=0$ and do show a finite threshold. Although a finite value of $D_t$ can be
generated by interactions even if $D_t=0$ at the single particle level in the
numerical model, this discrepancy suggests that other mechanisms not yet
understood may be at play here.

For finite $D_\text{r}$ one must in general retain the dynamics of the polarization.
The analysis of the modes in this case can be found in Ref.~\cite{fily:14} and
yields a growth rate with the characteristic behavior expected for a spinodal,
with a maximum growth rate  at wavevector $\sim\sqrt{-{\cal D}/2KD_t}$.

By including noise in the continuum model, one can examine the spectrum of
fluctuations and evaluate the structure factor
$S(q)=\langle\delta\rho_{\bm{q}}\delta\rho_{-\bm{q}}\rangle$. At small
wavevector one finds $S(q)\sim 1/(q^2+\xi^{-2})$, with
$\xi=\sqrt{KD_t/(D_\text{r}{\cal D})}\sim[\rho-\rho_c(\Pe)]^{-1/2}$ a
correlation length that diverges on the spinodal line~\cite{fily:12}.  Both the
divergence of $S(q=0,\rho)$ at $\rho=\rho_c$ and the scaling $S(q,\rho_c)\sim
q^{-2}$ are consistent with the behavior at an equilibrium gas-liquid spinodal
line.  Although the numerics can be fitted to this behavior~\cite{fily:12}, the
divergence of $\xi$ is generally masked by nucleation processes that further
support the liquid-gas analogy~\cite{redner:13}.  A true divergence is expected
at the critical point at $\Pe^\text{c}$, but it has not yet been possible to
extract numerical values for the critical exponents.
It has also been shown numerically that the coarsening dynamics closely resembles that of an
equilibrium spinodal decomposition and that activity  weakly affects on the
growth in time of the domains size $L(t)$ that is found to be very close to the
Cahn-Hilliard form, $L(t)\sim t^{1/3}$, describing the coarsening of a passive
system with no momentum conservation\footnote{Values of the coarsening
exponent slightly smaller that $1/3$ have been reported in
simulations~\cite{redner:13,stenhammar:13}, but it is not clear whether these
simulations had indeed reached the asymptotic regime.} although differences
exist when gradient terms accounting for interfacial phenomena are included in
the theory~\cite{redner:13,stenhammar:13,wittkowski:14}.  In contrast,
experiments in bacteria and active colloids generally do not see complete
separation in two bulk phases, but rather observe the assembly of finite-size
clusters that persist for very long times. In bacterial suspensions birth/death
events that eliminate density conservation have been shown to arrest the phase
separation yielding the formation of finite-size
clusters~\cite{cates:10,yang:14}. Attractive interactions have also ben
suggested as a possible mechanism for arresting the  phase
separation~\cite{palacci:13}, but the mechanism that control the size of
finite clusters in active colloids remain to be understood.

The $q^{-2}$ divergence of the structure factor as $q\rightarrow 0$ is also the
signature of the giant number fluctuations  (GNF) predicted and observed ubiquitously
in the ordered state of flocking
models~\cite{toner:95,simha:02,ramaswamy:03,marchetti:13}. MIPS and GNF
are, however, two distinct phenomena. In repulsive SPPs  the large density fluctuations occur
in a disordered state and are associated with a critical point and phase
separation. The large density fluctuations that occur in the ordered state of
active fluids with both polar and nematic  order are the consequence of the
coupling of a conserved field - the density - to a non-conserved noisy
polarization field.  The $q^{-2}$  behavior of the correlation is in fact a
generic feature of systems where the deterministic dynamics of a conserved
field couples to non-conserved noise, resulting in power-law decay of
correlations~\cite{grinstein:90}, and is also ubiquitous in driven granular
materials with inelastic interactions~\cite{noije:99}.

The onset of MIPS suggests that at moderate density a system of repulsive
active colloids behaves like a thermal system with attractive interactions.
This correspondence was recently demonstrated by correlating the pressure measured  in sedimentation experiments of active colloids with numerical simulations~\cite{ginot:15}. 
It has additionally been explored in models of Brownian particles with Gaussian colored noise correlated over a time $\tau_\text{r}$, where
 the colored nature of the noise can be incorporated as an effective
interaction in a many-body Fokker-Planck equation~\cite{fox:86,fox:86b}. Using this
approach it has been shown recently that for weakly persistent motion (small
$\tau_\text{r}$) activity can be recast in the form of an effective pair potential
with an attractive part of strength that increases with increasing
$\Pe$~\cite{farage:15}. On the other hand, when the pair interparticle
potential actually has an attractive component,  activity can yield a weak repulsive contribution to the effective potential, hence competing with
attraction by promoting particle escape from the potential well. In other words,
self-propulsion can actually drive a system that has undergone bulk phase
separation due to attractive interactions back into a uniform state.  It should be noted, however, that the ABP model and the model with Gaussian colored noise are characterized by different speed distributions, which is Gaussian in the latter, but bounded by $\vel_0$ in ABP and in run-and-tumble models. Simulations have additionally demonstrated that the
competition of attraction and activity yields an intermediate regime where
complete phase separation is arrested and particles assemble in a gel-like
structure that has been seen in simulations in both two and three
dimensions~\cite{redner:13a,farage:15, prymidis:15}

Finally, it has been shown that hydrodynamic interactions can suppress phase
separation by aligning the particles and therefore suppressing the
self-trapping responsible for cluster formation~\cite{matas-navarro:14,zottl:14}.
On the other hand, strong clustering is often observed in active systems even
in the presence of aligning rules~\cite{marchetti:13}, indicating that more work
needs to be done to understand the effect of aligning torques on the MIPS of
repulsive active particles.

\section{Active crystals and active glasses}
\label{crystals}

At large packing fraction ABP have been shown to freeze in solid
states. While this may not seem surprising since purely repulsive systems are
known to form crystalline and glassy states in equilibrium, it is not a priori obvious
that such states can survive in the presence of continuous energy input at the
microscopic scale. Key to the formation of active solid states is the fact
that steric repulsion  strongly suppresses motility at high density, as discussed in
Section \ref{MIPS}. This results  in particle caging and in what was first
described by Henkes and collaborators as `active jamming'~\cite{henkes:11},
although that work was for  a model of repulsive SPPs with alignment
interactions, where confinement was required to impede collective flocking. 

In equilibrium freezing in two dimensions is a continuous, two-step process,
where the system goes from a liquid to a hexatic liquid state with quasi-long
range bond orientational order, and then to a crystalline state with long-range
orientational order and quasi-long range translational order.  Crystallization
can be quantified using both structural and dynamical criteria. The onset of
translational order identified via structural probes generally coincides with
criteria for dynamical arrest, such as the vanishing of the long-time diffusion
coefficient $D=\lim_{r\rightarrow\infty}
\frac{1}{4t}\langle\Delta\bm{r}(t)]^2\rangle$.  Crystalline order has been seen
in phase separating active systems both in experiments and simulations~\cite{bialke:13,palacci:13,reichhardt:14,menzel:14}. In
simulations of monodisperse systems the dense phase in the phase separated
regime exhibits substantial structural order resembling a colloidal crystal
near the hexatic-crystal transition, but with unusual super-diffusive
behavior~\cite{redner:13}. Crystalline clusters have also been seen in the
experiments, such as the `living crystals' formed by light activated colloids
engineered by Palacci \emph{et al.}~\cite{palacci:13}.  Numerical simulations have also examined the
onset of bond-orientational order in ABP~\cite{redner:13} and its correlation with
structural arrest, concluding that the onset of order remains continuous, but
is shifted to higher density that in equilibrium systems~\cite{bialke:12}. Additionally, Bialke
\emph{et al.}~\cite{bialke:12} reported the existence of an intermediate
region between the liquid and the solid states characterized by large
structural fluctuations. It is tempting to identify this phase an an active
hexatic phase similar to but distinct from the equilibrium hexatic that has
been observed for instance in superparamagnetic colloidal particles confined to
an air-water interface~\cite{zahn:00}. Finally, in models with aligning
interactions the interplay of activity, alignment and structural order can lead
to  partially ordered liquid crystalline
resting and traveling states  that are beyond the scope of the present article~\cite{chen:15,adhyapak:13}.

\begin{figure}[h!]
\centering
\includegraphics[width=0.99\linewidth]{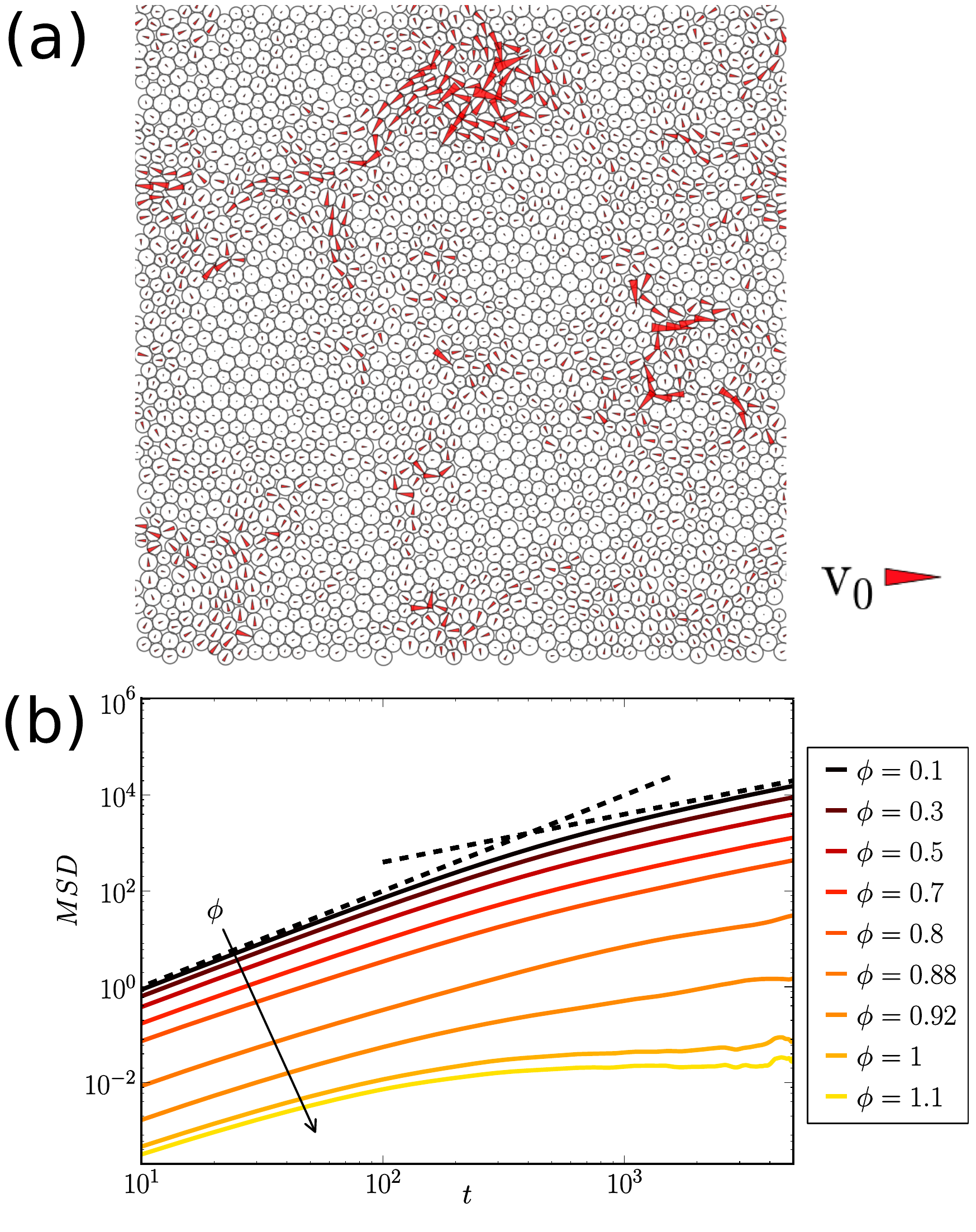} 
\caption{{(a) Snapshot of a configuration near the active glass transition  from our simulations of polydisperse ABP,  reproduced from \cite{fily:14}, arrows are velocity vectors. Note the similarity to dynamical heterogeneities and low energy disordered modes. (b)}
MSD versus time for various packing fractions $\phi$ from our simulations of
polydisperse ABP with soft repulsive interactions,
Eqs.~(\ref{eq:ri}--\ref{eq:thetai}), for $\Pe=20$ showing ballistic growth at small
times and the crossover from diffusive to bounded behavior with increasing
packing fraction at long time. The dashed lines have slopes $2$ and $1$. 
}
\label{fig:MSD}
\end{figure}

Our own work has focused on polydisperse systems, where even in equilibrium
crystalline order is bypassed in favor of glassy or jammed states. {Fig.~\ref{fig:MSD}(a) shows a simulation snapshot near the active glass transition for a non-aligning, polydisperse system.} In this case
the characterization of  glassy behavior relies on criteria for dynamical
arrest, such as the strong suppression of diffusion at intermediate times, when
particles are persistently caged. Fig.~\ref{fig:MSD}(b) shows the evolution of the
mean square displacement  of a polydisperse assembly of repulsive SPPs over a
range of packing fractions. At small and intermediate packing fraction the
long-time dynamics remains diffusive, but at high packing fraction the MSD is
bounded over the time of the simulation, indicating a jammed or frozen state.
The fluctuations in the actively jammed state, {see Fig.~\ref{fig:MSD}(a)}, are intimately connected to the
low frequency or `soft' modes of a solid that are associated with the jamming
transition of passive thermal systems~\cite{henkes:11}. At small $\vel_0$
 active systems form solid phases that can be analyzed in terms of a
complete set of normal modes with normal frequencies $\omega_\nu$ determined by
the inter-particle interactions~\cite{henkes:11}.  As in the familiar description of lattice
vibrations in terms of phonons, the lowest frequencies correspond to large
scale collective structural rearrangements of the system. Work on passive
granular matter has shown that such low frequency modes have much lower energy
barriers than high frequency modes~\cite{silbert:05}. The jammed state is
then a steady state with structural rearrangements involving a  handful of low
frequency modes that also provide easy paths for unjamming as $\vel_0$ is
increased. This hints at a generic connection between the dynamics of jammed
active systems at low $D_\text{r}$ and the kinetic arrest of passive glassy systems,
as explored in Refs.~\cite{berthier:13,berthier-szamel:15,farage:15}. Note that
the prominence of low frequency modes in the active jamming dynamics is
strongly linked to the polarization - velocity alignment mechanism, and more
work is necessary to disentangle the interplay of activity, alignment and
caging.
%

One interesting feature of the onset of glassy behavior in repulsive SPPs is
the shift of the glass transition to packing fraction higher than the value
$\phi_G\approx 0.8$ obtained for thermal systems in the limit of vanishing
motility, $\vel_0\rightarrow 0$~\cite{berthier:14,ni:13,fily:14}. This shift
increases with increasing persistence time $\tau_\text{r}$ and has been reported in
both simulations of SPPs with hard repulsions where activity pushes the glass
transition towards the random close packed limit of $\phi_{RCP}=0.842$ and in
our simulations of soft disks (see phase diagram of Fig.~\ref{fig:PD}), where
we find $\phi_G(\vel_0\rightarrow 0)\approx \phi_{RCP}$ for $D_\text{r}/\mu k=5\times 10^{-4}$.

Finally, Ni \emph{et al. }have shown numerically  that doping a hard sphere
glass with a small number of active particles promotes crystallization
bypassing the glass transition~\cite{ni:14}. This suggests that a small
concentration of active particles may be used as a means for facilitating the
formation of large crystalline states that would be otherwise hindered by
defects and grain boundaries.

\section{Discussion and Outlook}\label{discussion}

Collections of repulsive self-propelled particles, also known as active Brownian particles, provide a minimal realization of an active system with rich non-equilibrium behavior. They exhibit active gas, liquid and solid phases with novel mechanical properties. 
At low density they {seemingly} behave as `hot' colloids with an effective temperature controlled by activity {(but see Ref.~\cite{solon:15c})} and a novel swim pressure unique to active systems.  At intermediate density they resemble attractive colloids and exhibit a motility-induced phase separation. At high density they form solid and glassy phases that have provided new insights in the physics of glassy and jammed solids. 
{
Mapping ABP's onto an ``equivalent'' equilibrium system has proven a powerful method, as evidenced by the concepts of effective temperature and effective attraction. 
Such mappings, however, are necessarily limited in scope.
Activity-induced attraction is antagonistic to actual attraction~\cite{redner:13a}. 
The hot colloid model fails in the presence of a rapidly varying external potential, e.g., a stiff wall~\cite{solon:15b,solon:15c}. 
As a result, the application of these concepts is sometimes subject to controversy and one should keep in mind that active systems are fundamentally nonequilibrium. 
}

The key property that has allowed much theoretical progress in the description
of this minimal active system is that the noisy single-particle angular
dynamics is decoupled from interactions. In other words active Brownian
colloids do not exert torques on each other nor on the walls of a container. As
a consequence the effect of activity can be mapped onto a non-Markovian noise
of strength controlled by the self-propulsion speed and with time correlation
given by the persistence time $\tau_\text{r}$, as shown in Eqs.~(\ref{eq:ri-one}--\ref{eq:xi}). This has led to renewed interest in stochastic models of Brownian particles with Gaussian colored noise, for which exact results exist in the literature~\cite{fox:86,fox:86b,jung:87,maggi:15}. Although these models are not precisely equivalent to ABP due to the different speed distributions, useful insight has been obtained from these studies~\cite{farage:15,berthier-szamel:15}. On the other hand, torques are likely to be
present in most experimental realizations, as they can arise from the
non-spherical shape of the particles, from lateral friction, from hydrodynamic
interactions or explicit aligning rules, and will generally modify the
collective behavior of the systems. It has already been shown, for instance,
 that the pressure  is a state function only for the case of repulsive spherical active Brownian particles, independent of the nature of the interaction of
the system with the walls. The presence of torques may also arrest the bulk
phase separation, leading instead  to clustering or micro-phase separation, as
ubiquitously seen in experiments. Work remains to be done to fully understand
how alignment, induced either by anisotropic steric effects or by aligning interactions,
modifies the phase behavior of active colloids and to make \emph{quantitative} contact
with experiments. 

In the limit of small persistence time ($\tau_\text{r}\rightarrow 0$), active Brownian colloids are equivalent to Brownian colloids. The opposite limit of $\tau_\text{r}\rightarrow \infty$   is  intriguing and has not been fully
explored. In this limit the active drive becomes correlated in time and
resembles quenched disorder. At high density the fluctuations patterns tend to exhibit both
spatially and temporally long-ranged correlations~\cite{fily:unpub} and
resemble those seen in sheared passive granular matter~\cite{vagberg:11,bi:15},
suggesting an intriguing connection that
needs to be further investigated. 

Finally, quite unexplored is the effect of noise in the continuum dynamics as described for instance by Eqs.~(\ref{rho}) and (\ref{p}). Most work using phenomenological continuum models to describe active matter has focused on deterministic equations or has included white Gaussian noise in the linear theory to evaluate correlation functions~\cite{fily:12}. On the other hand, the noise in active systems is in general  multiplicative in the density~\cite{bertin:13} and may, for instance, provide a mechanisms for 
selecting nonlinear dynamical states.  The minimal model of active Brownian colloids may provide an excellent playground for beginning the exploration of  the role of multiplicative noise in mesoscopic theories of active matter.

\section{Acknowledgments}
MCM thanks Xingbo Yang  and Lisa Manning for their contribution to some aspects of the work reviewed here and for fruitful discussions. MCM was supported by NSF-DMR-305184. MCM and AP acknowledge support by the NSF IGERT program through award  NSF-DGE-1068780. MCM, AP and DY  were additionally supported by the Soft Matter Program at Syracuse University. AP acknowledges use of the Syracuse University HTC Campus Grid which is supported by NSF award ACI-1341006. 
YF was supported by NSF grant DMR-1149266 and the Brandeis Center for Bioinspired Soft Materials, an NSF MRSEC, DMR-1420382.





\section*{References\footnote{Several references of special relevance have been
marked with an (*).}}

\end{document}